\documentclass[prb,aps,preprint,epsfig]{revtex4-2}
\usepackage{CJK}
\usepackage{graphicx}% Include figure files
\usepackage{dcolumn}% Align table columns on decimal point
\usepackage{bm}% bold math
\usepackage[utf8]{inputenc}
\usepackage{xcolor}
\bibpunct{(}{)}{,}{s}{,}{,}
\makeatletter

\newcommand{\Rmnum}[1]{\expandafter\@slowromancap\romannumeral #1@}
\makeatother

\begin{document}
\begin{CJK*}{GB}{} % Use default fonts from CJK (see below)
\title {Probing Metal-Molecule Contact at the Atomic Scale via Conductance Jump}
\author{Biswajit Pabi, Debayan Mondal, Priya Mahadevan*, and Atindra Nath Pal* }
\vspace{1.5cm}
\address{Department of Condensed Matter Physics and Material Science, S. N. Bose National Center for Basic Science, Sector III, Block JD, Salt Lake, Kolkata - 700106 }
\vspace{1.5cm}
\address{Email: atin@bose.res.in, priya@bose.res.in }

%\date{\today}
\begin{abstract}
Understanding the formation of metal-molecule contact at the microscopic level is the key towards controlling and manipulating atomic scale devices.
Employing two isomers of bipyridine, $4, 4^\prime$ bipyridine and $2, 2^\prime$
bipyridine between gold electrodes, here, we investigate the formation of metal-molecule bond by studying charge transport through single molecular junctions using a mechanically controlled break junction technique at room temperature. While both molecules form molecular junctions during the breaking process, closing 
traces show the formation of molecular junctions unambiguously for $4, 4^\prime$ bipyridine via a conductance jump from  the tunneling regime, referred as `jump to molecular contact', being absent for $2, 2^\prime$
bipyridine. Through statistical analysis of the data, along with, molecular dynamics and first-principles 
calculations, we establish that contact formation is strongly connected with the molecular structure of the electrodes as well as how the junction is broken during breaking process, providing important insights for using a single-molecule in an electronic device.

\vspace{10 mm}

\end{abstract}

%/PACS:  75.47.Gk,  73.40.-c,  75.70.-i

\maketitle
\end{CJK*}

\section{Introduction}
The drive towards miniaturization has led to envisaged electronic components having elements which
have dimensions in the nanometre scale. These elements could be molecules~\citep{aviram_Ratner}, and before placing them
in a circuit, it becomes essential to know how they would behave. While there have been tremendous
advances made~\citep{diode_Latha,Elke_MR,Reddy_thermal,Atin_Natcom}, probing the contact formation at the microscopic level has not been
straightforward. A route towards exploring atomic scale transport has involved forming mechanical break junctions~\citep{Ruitenbeek_PRB_oneatompointcontact}.
Considering metallic electrodes, examining the metal-metal contact formed in a break junction at
liquid helium temperatures, one finds a regime in which one has quantum tunneling, followed by the
formation of contacts. This transition from a tunneling transport regime to contact formation occurs, in most cases, 
via a sudden jump in the conductance, referred to as a jump to contact~\citep{Ruitenbeek_PRL_JUmptocontact,PRB_Wulfheckel_2020}. Various metals like
Au, Pt, Cu and Ag exhibit a jump to contact, while there are others like Ir, Ni and W in which a
jump to contact can be absent. In a controlled experiment
involving an Au break junction, using a highly ordered gold electrode, it was shown that the
phenomena could be explained by a generic potential energy model with the elastic constant of the
metal being the only free parameter~\citep{Molenwees_PRL_JCJOC}.

Formation of metal-molecule bond is rather complex and highly dependent on the nature of binding groups and geometry of the
electrodes~\citep{Latha_JACS_contact_linker,Latha_Nanolett_2006_linker,PRL_linker_Latha,Tao_linker_2006_JACS,linker_Pera_2010,kiguchi_pccp_2017,Ran_Nanoletter_2014,Atin_BJNano}. Jump to molecular contact was reported for several flat molecules through scanning tunneling microscopy (STM) based experiments at cryopgenic temperatures~\citep{haiss2006_NatureMat,Wulfhekel_2008_PRB,Wulfhekel_2011_Nanturenano,Wulfhekel_2012_Nanolett}. It is believed that there exist two energetically close adsorption geometries with one of them bridging
the junction via a soft phonon from the molecular side groups~\citep{Wulfhekel_2008_PRB}. In case of mechanically controllable break junctions, however, the shape of the
electrodes are not expected to remain in line while approaching each other, especially at room
temperature~\citep{Kiguichi_PCCP_Breakingandmaking}. This may lead to the formation of asymmetric
junctions, which, in contrast to the breaking process, may not provide precise conductance of single molecular junctions. In recent break junction experiments with an organometallic compound (BdNC), carried
out in solution, a jump was observed in the conductance for closing
traces~\citep{Calame_NatCom_Insituformation}. This was attributed to the formation of
one-dimensional coordination polymers in those junctions. The knowledge about the role of metal-molecule interaction towards bond formation is rather limited and a general picture is still lacking.

In the present study, we report the observation of jump to molecular contact in a single molecular junction by studying the charge transport using a mechanically controllable break junction setup at ambient
temperature. Through statistical analysis of conductance-distance traces, along with density functional theory based calculations and ab-initio molecular
dynamics, we provide the microscopic origin behind the metal-molecular contact formation. 
Considering two isomers of bipyridine, $4, 4^{\prime}$ bipyridene ($4,4^{\prime}$-BPY) and $2, 2^{\prime}$ bipyridene ($2,2^{\prime}$-BPY), attached to gold electrodes, we find that molecular junctions form in the breaking traces for both the molecules. The surprising aspect is that $4,4^{\prime}$-BPY forms molecular junctions isotropically with conductance jumps in the closing traces also, an aspect that is absent for $2,2^{\prime}$-BPY.Our DFT and MD simulations reveal two important mechanisms for jump to molecular contact. Firstly, while the $2,2^{\prime}$-BPY prefers to lie flat on the Au surface, $4,4^{\prime}$-BPY has
two stable minima - one with the molecule lying flat on the surface and the other with the molecule standing vertically, despite having similar anchoring group involving nitrogen. This unusual aspect of the $4,4^{\prime}$-BPY emerges from the fact that the nitrogen forms a very strong bond in $4,4^{\prime}$-BPY~\citep{Hou_2005} while the orientation on the Au surface in $2,2^{\prime}$-BPY doesn't allow a similar strong bond due to steric effects from the hydrogens attached to the carbon atoms, with the Au-N bondlength equal to 2.24 $\AA$ here in contrast to 2.12 $\AA$ for the $4,4^{\prime}$-BPY molecule. The possibility of having two stable energy conformations may lead to a jump to contact for $4,4^{\prime}$-BPY, similar to Ref[\citep{Wulfhekel_2008_PRB}].  Additionally, we propose another mechanism, according to which, the observed jump to molecular contact in case of $4,4^{\prime}$-BPY may arise from the fact that the molecule breaks by pulling few gold atoms with it due to strong Au-N bond compared to the Au-Au bond. Consequently, while making the contacts during the closing loops, one does not have to worry about the directionality of the bonding. Through a statistical analysis of the experimental data the second mechanism was found to be the dominant one.

\begin{figure*}[!t]
\begin{center}
\includegraphics[width=1\linewidth]{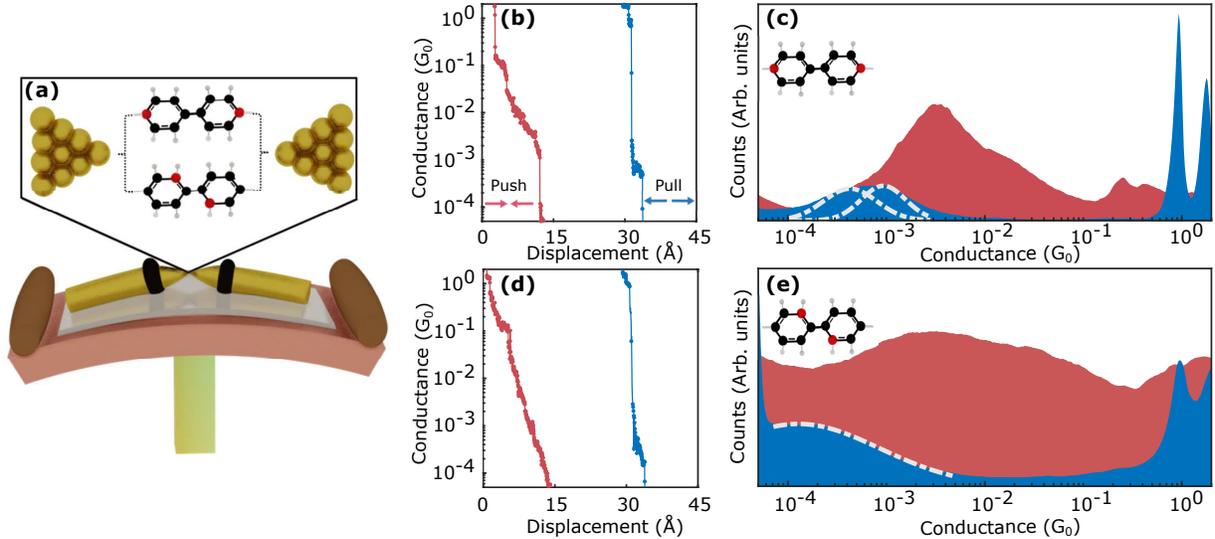}
\end{center}
\caption{ (a) Schematic of a mechanically controllable break junction
(MCBJ) set up, along with the schematic representation of the molecules used
in this work: $4, 4^\prime$ bipyridine ($4,4^{\prime}$-BPY) and $2, 2^\prime$ bipyridine ($2,2^{\prime}$-BPY).
Conductance traces for (b) $4,4^{\prime}$-BPY and (d) $2,2^{\prime}$-BPY. Blue (red) represents trace
during pull (push). 1D logarithmic conductance histogram for (c) $4,4^{\prime}$-BPY and
(e) $2,2^{\prime}$-BPY, constructed from 5000 and 8000 traces respectively, without any
data selection. Blue and red represent the corresponding pull and push
respectively. White dashed line shows the Gaussian fit to the corresponding
molecular peaks of the histogram. } \label{fig1}
\end{figure*}

\section{Methods}

A single molecular junction was formed in a custom built mechanically controllable break junction
(MCBJ) technique at ambient condition~\citep{Muller_PRL_MCBJ}. A gold wire (0.1~mm dia, 99.998$\%$,
Alfa aesar) with a notch at its center was fixed onto a flexible substrate (Phosphor Bronze) with a
two component epoxy glue (Stycast 2850 FT with catalyst 9). For creating molecular
junctions, molecules ($4,4^{\prime}$-BPY and $2,2^{\prime}$-BPY, 98$\%$,~Alfa aesar) were either
evaporated or drop casted (0.1mM Ethanol solution) on top of the notch and the wire was controllably broken using a piezo stack. Formation of molecular junctions were confirmed from the appearance of additional peaks in the conductance histograms below 1G$_0$
(where G$_0 = 2e^2/h$, is the quantum of conductance). In the conductance-distance measurements, the
typical moving rate of the piezo was 3.2 $\mu$m/s, which translates into an opening and respective
closing rate for the two leads of about $2-4$ nm/s (attenuation factor of $\sim 10^{-3}$)~\citep{PRB_vanwees_mechanics}. 

In order to model the experiments, we use density functional theory based ab-intio electronic structure calculations  using the generalised gradient approximation (Perdew Burke-Ernzerhof)\cite{PhysRevLett.77.3865} for the exchange correlational functional with dispersion-corrections(DFT-D2)~\citep{grimme2006semiempirical,grimme2004accurate} as implemented within Vienna ab initio simulation package (VASP) ~\citep{Kresse1996a,Kresse1996,Kresse1994,Kresse1993}. We have used $\Gamma$ point K-space sampling \cite{monkhorst1976special} and a plane-wave cut-off of 500 eV . The structural relaxations were carried out till forces were less than $10^{-3}$ eV/atom and energy convergence in each case was $10^{-5}$ eV.
The ab-initio MD simulations ~\citep{frenkel2001understanding,allen2017computer,kresse1993ab} were carried out at room temperature (300K) and the whole system is kept at the desired temperature in NVT canonical ensemble using a Nos\'{e}-Hoover thermostat~\citep{nose1984unified,shuichi1991constant,hoover1985canonical} for several molecular junctions formed by $4,4^{\prime}$-BPY and $2,2^{\prime}$-BPY molecules with gold electrodes. The optimized structure of a gold slab-molecule-gold slab is used to construct the junction atomic geometry, where 8 layers were used to construct the slab.  The slabs are separated by a vacuum of 15 $\AA$. The electrodes are pulled apart in steps of 0.1$\AA$. The entire system is relaxed at every step by taking 200 MD steps of 0.5 fs.

\section{Results and Discussion}
Typical opening or pull (blue) and closing or push (red)
conductance traces for both the molecules are shown in Figure~1b and~1d
respectively (see SI for measurement details). During junction breaking process (pull), well-developed
molecular plateaus in the range of $10^{-3}$ to $10^{-4}$ G$_0$ is noticed
for both $4,4^{\prime}$-BPY $\&$ $2,2^{\prime}$-BPY.  Closing traces (push) of $4,4^{\prime}$-BPY, however, exhibit
a plateau like feature which is longer and slanted compared to the pull
traces, in line with the previous observations~\citep{WenzingHong_Cellpress_tunnelingthroughspace}. In case of $2,2^{\prime}$-BPY, no such
plateau was observed in push traces, instead, a continuous tunneling-like 
behavior observed. A closer look at the push traces for $4,4^{\prime}$-BPY reveals that
atomic contacts of Au forms via two jumps in the conductance: first jump
occurring from the background noise or tunneling to a conductance value of
$\sim$ $10^{-3}$ G$_0$ and the second one from $\sim$ $0.1$ G$_0$ to the 
few tenths of G$_0$.  First jump is referred as `jump to molecular contact'
(abbreviated as J2C) and the second jump corresponds to the formation of
metallic contact. No such jump is observed in case of $2,2^{\prime}$-BPY. In the following
paragraphs we discuss more details about origin of the J2C phenomena.

\begin{figure*}[!t]
\begin{center}
\includegraphics [width=1\linewidth]{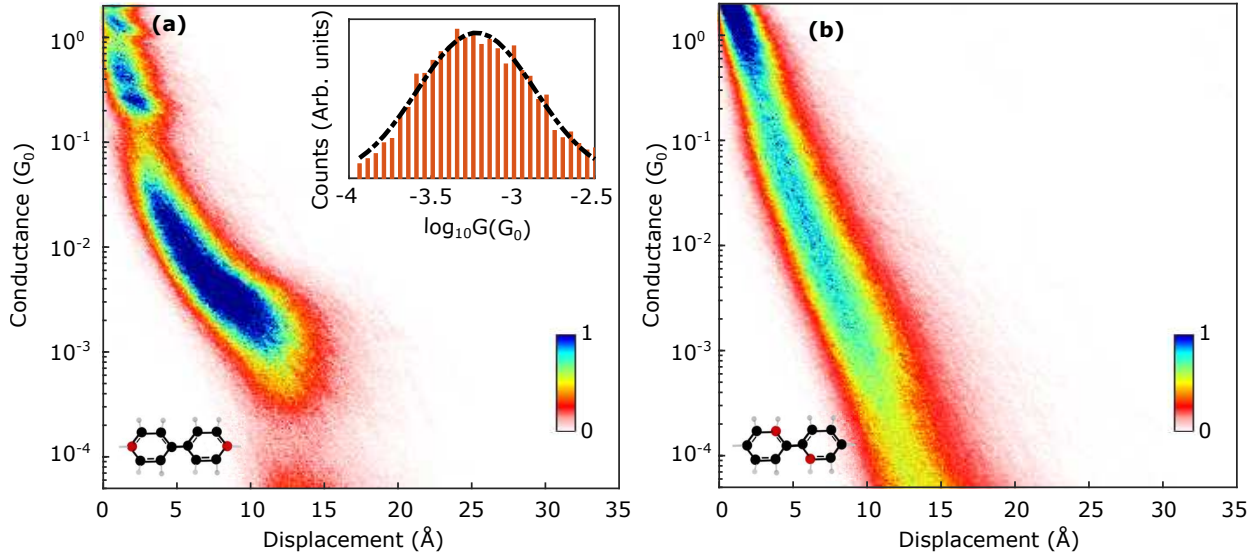}
\end{center}
\caption{ 2D Conductance-displacement density plot for (a) $4,4^{\prime}$-BPY and (b) $2,2^{\prime}$-BPY, for push traces used in Figure~1d and 1e, with  100 bins per decade. Inset: Histogram of the conductance of the molecular contact formed after the jump, having 3750 traces (see SI for details). } \label{fig2}
\end{figure*}

Thousands of individual traces were collected, analyzed and plotted as histograms to get a meaningful estimate of the molecular conductance from statistically independent
junction configurations. Figure 1c and 1e display the logarithmically binned normalized
one-dimensional (1D) conductance histogram for $4,4^{\prime}$-BPY and $2,2^{\prime}$-BPY
respectively, constructed from 5,000 and 8,000 conductance traces without any data selection. For
both the molecules, the pull histogram (blue) exhibits a clear peak around 1G$_0$ corresponding to
the Au atomic contact, followed by a prominent molecular conductance feature in the range of $5
\times 10^{-2}$ G$_0$ to $5 \times 10^{-4}$ G$_0$ ($4,4^{\prime}$-BPY) and $1 \times 10^{-3}$ G$_0$
to $1 \times 10^{-4}$ G$_0$ ($2,2^{\prime}$-BPY). For $4,4^{\prime}$-BPY, conductance peak can be
resolved into two separate peaks ($9.33\pm 0. 29 \times 10^{-4}$ G$_0$ and $4.28\pm 0.07 \times
10^{-4}$ G$_0$), obtained by fitting Gaussian to the characteristic maxima, similar to the previous observation~\citep{Latha_naturenan0_BioyridineSwitch, Kiguichi_JACS} (see SI Figure~S5
for details). The most probable conductance for $2,2^{\prime}$-BPY was found to be $1.27\pm 0. 05 \times 10^{-4}$ G$_0$ (Figure~1e). Push histogram (Red) for $4,4^{\prime}$-BPY (Figure~1c) shows a clear conductance peak
in the region ($5\times 10^{-4}$ G$_0$ to $2 \times 10^{-1}$ G$_0$), being higher than the pull
conductance, whereas, no such conductance peak is observed in case of $2,2^{\prime}$-BPY.

To get a clear picture about the evolution of conductance during the
formation processes and understand the possible origin of observed J2C in
case of $4,4^{\prime}$-BPY, we have constructed conductance-displacement histogram from
the push traces for both the molecules. Figure~2a and ~2b display such 2D histograms for the push data for $4,4^{\prime}$-BPY and $2,2^{\prime}$-BPY, respectively. The plots
were constructed from the same  traces used in the 1D histogram, by aligning them at 2G$_0$
and having 100 bins per decade. Corresponding pull histograms are shown in SI Figure~S2. By analyzing the length histogram, we see that $4,4^{\prime}$-BPY (Figure~2a) exhibits a
longer and slanted conductance plateau in push with an average length of $9.77\pm0.05$$\AA$ (SI Figure~S3c), compared to the  corresponding pull plateaus with an average plateau length of $3.03\pm0.02$$\AA$ (SI Figure~S3a). Similar analysis for $2,2^{\prime}$-BPY (pull) shows an average plateau length of $2.57\pm0.01$$\AA$ (SI Figure~S3b). The most intriguing aspect is the unambiguous formation of
single molecular junction for $4,4^{\prime}$-BPY during the closing process with a conductance-jump occuring mostly at $\sim (5.98\pm0.23)\times 10^{-4}$G$_0$, calculated from the Gaussian fit of the histogram of conductance values where jump to molecular contact takes place, shown in the inset of Figure~2a (see SI for details about the jump detection algorithm). This value bears close resemblance to the conductance of of Au-$4,4^{\prime}$-BPY-Au junctions in the stretched configuration, reported earlier~\citep{Latha_naturenan0_BioyridineSwitch}. Jump to molecular contact is clearly absent for $2,2^{\prime}$-BPY (Figure~2b), where a continuous tunneling slope is observed from the noise background while closing the electrodes and
no clear formation of molecular junction was noticed. 

\begin{figure}
\begin{center}
\includegraphics [width=0.75\linewidth]{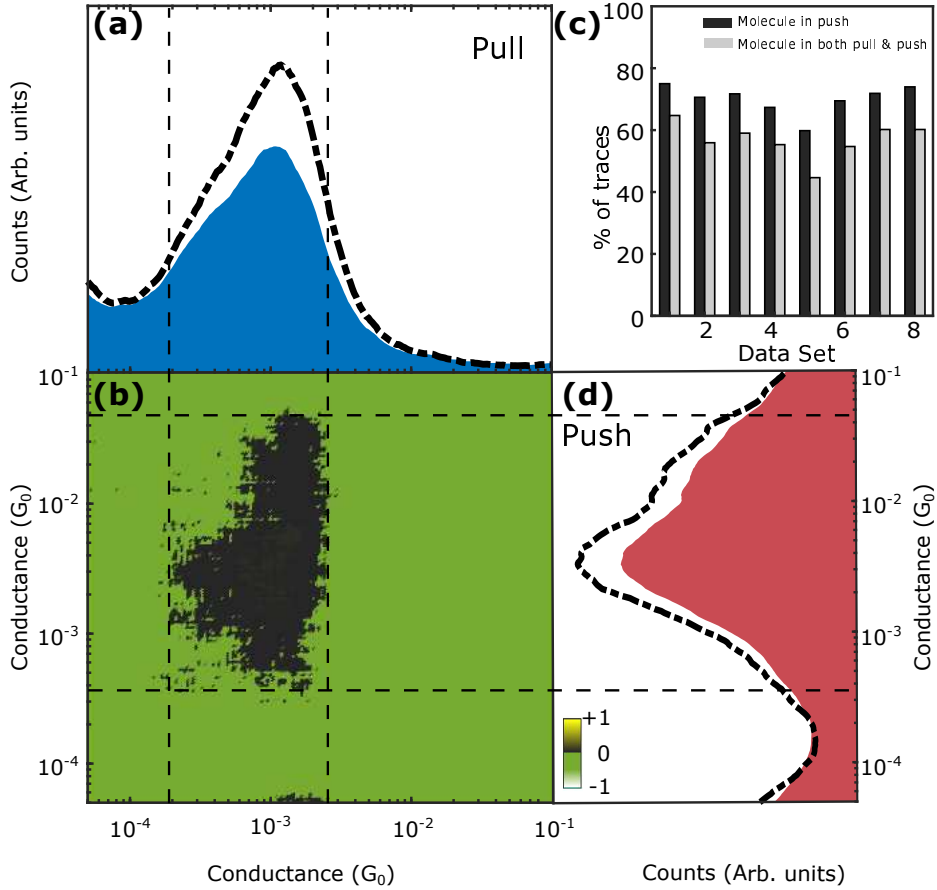}
\end{center}
\caption{ Conditional histogram of $4,4^{\prime}$-BPY for (a) pull and (d) push traces.  Black dashed-dotted line represent the histogram of the selected pull (push) traces for which plateaus are formed in the corresponding push (pull) traces, respectively. Blue and red area graph represents the histogram for all traces, as a reference for (a) and (d), respectively. (b) 2D correlation histogram, indicating the correlation between the pull (horizontal axis) and push (vertical axis ). Histogram is constructed from 10000 traces without any data selection using 50 bins per decade. (c) Bar diagram showing percentage of traces forming molecular junction in push via jump (black) and percentage of traces having both molecule in the pull and push (grey) for 8 different data set.  } \label{fig3}
\end{figure}

A statistical correlation analysis of the conductances traces, as described in
Ref~\citep{Halbitterlatha_AcsNano_Correlation}, has been used to distinguish various structural
configurations and understand their origin. Using the same procedure, we have carried out
statistical correlation analysis between the pull and push
traces~\citep{Halbitter_Nanoscale_pull-pushcorrelation} to find out if there is any possible
correlation between them (see SI for details). Figure~3b shows the
two-dimensional cross correlation plot for $4,4^{\prime}$-BPY, constructed from 10,000 traces with
50 logarithmic bins per decade. The vertical axis belongs to the push traces and the horizontal to
the corresponding pull traces. A clear positive correlation between the pull and corresponding push
traces is observed, shown by a dark spot near the region where pull and push peak appear, as
depicted by the black dashed lines.This is also evident from the conditional histogram of the corresponding pull (push) traces, shown by the black dash-dotted line in Figure~3a (3d). While selecting the push (pull) traces having conductance plateau in the selected region between the black dashed lines,  we observe a clear enhancement in the conductance peak of the pull (push) histogram. A positive correlation may indicate a structural memory
effect~\citep{Halbitter_JCP_StructureMemory} that any molecular junction in push is possible, mostly
when a molecular junction is formed during breaking process (pull). Similar positive correlation
between pull and push traces was observed for Au-CO-Au junction at cryogenic
temperature~\citep{Halbitter_BJNano_CorrelationofAu-Co-Au}. It was concluded that the molecule is
bound rigidly to the apex of one electrode and the similar single molecular configuration was
reestablished during closing process. At room temperature, however, due to the enhanced surface
diffusion, electrodes are flattened and a correlation between the opening and closing traces is
uncommon. Figure~3c compares the correlation observed in push from eight different data sets
from three different samples with a varying probability of molecular junction formation. A
conditional analysis of eight statistically independent data sets indicate that J2C
was observed for more than $80\%$ cases whenever a molecular junction is formed during pull, establishing the fact
that J2C is primarily associated with the formation of a molecular junction in pull (see SI for details).

\begin{figure*}[!htb]
\begin{center}
\includegraphics [width=0.8\linewidth]{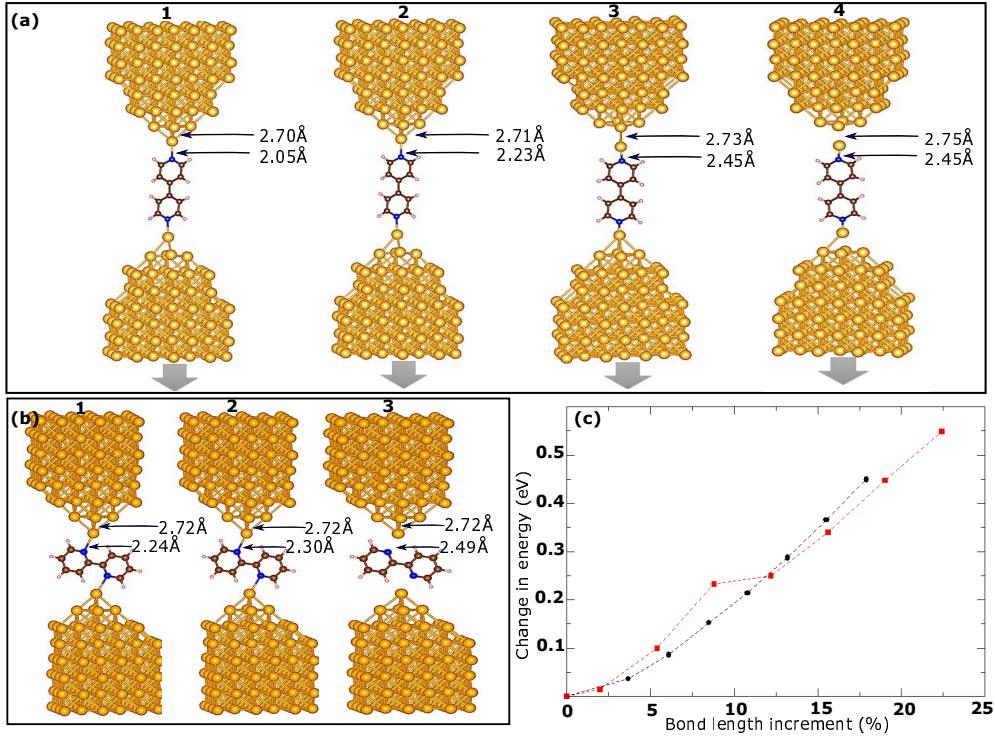}
\end{center}
\caption{ (a) MD snapshot of molecular junction with $4,4^{\prime}$-BPY for effective length of electrode (1) 11.10$\AA$ , (2) 11.60$\AA$ , (3) 12.20$\AA$ , and (4) 12.45$\AA$ (b) MD snapshot of molecular junction with $2,2^{\prime}$-BPY for effective length of electrode (1) 6.65$\AA$ , (2) 6.80~$\AA$, and (3) 6.89$\AA$ . (c) Change in total energy with respect to stretching of the bond of Molecule-Au (black) and Au-Au atom (red).   } \label{figure4}
\end{figure*}
 
To simulate the experiment, we take two
pyramidal gold electrodes and place the molecule $4,4^{\prime}$-BPY symmetrically 
between the electrodes.
The optimised Au-molecule bond length was found to be $2.12\AA$ in the static calculation. Placing the molecule 
at this distance from the electrode, led to an effective length  
(distance between two electrode tips) of 11.10$\AA$. The Au-Au bondlengths in
the electrodes ware set equal to the bond length of 2.65 $\AA$ found in the experimental structure.
Molecular dynamics simulations were then carried out at a temperature of 300 K. After 200 iterations, we find that
while the positions of the Au atoms in the electrode do not change, two atoms from the electrode come
closer to the molecule making shorter bonds. We have increased the effective length to several values as shown in Figure~4a and there are two aspects that emerge. An Au atom gets removed from the electrode
and moves towards the molecule. Additionally the molecule is no longer symmetrically placed with respect to
electrodes, and the distance of the molecule and one end of the electrode becomes shorter
(on one side the length is 2.45 $\AA$ and on the other it is 2.55 $\AA$). The converged effective
length is now 12.20 $\AA$ (Figure~4a (3)). If we increase the effective
length further (13.00 $\AA$), we still find that the Au atom moves towards the molecule (Figure 4a~(4)).
Similar calculation was done for $2,2^{\prime}$-BPY. The N-Au bondlength is found to be 2.24$\AA$ where we optimised the structure at 0K. When we increase the separation, the Au atom from the electrode is not detached as found for $4,4^{\prime}$-BPY ( Figure~4b).

In order to understand this further, we have examined the energy variations as we stretch the Au-Au bond 
of a metal-metal contact. The energy is found to vary continuously till a bond length of 3.20 $\AA$
(a stretch of 8.8\% in the bondlength) (Figure~4c). At this point there is a discontinuous change 
in the bondlength and a jump in the energy also. The energy variations now lie on a new curve and are again
continuous. As the hopping interaction strength between any two atoms varies inversely with their distance
according to a power law, this jump in bondlength would translate into a jump in the hopping interaction strength.
This then leads to a jump in the conductance, and is consistent with experiments carried out with Au electrodes 
in which a jump to contact is observed~\citep{Ruitenbeek_PRL_JUmptocontact, Molenwees_PRL_JCJOC, PRB_Wulfheckel_2020}. When we went on to examine the variations in the energy of the molecule  connected 
to a Au atom, on subjecting it to a similar elongation we found  a continuous variation (Figure~4c). 
This suggests that one should see only a monotonic variation in the conductivity in this case.
This is, however, contrary to experiments where one has a jump to contact.

\begin{figure*}

\includegraphics [width=0.9\linewidth]{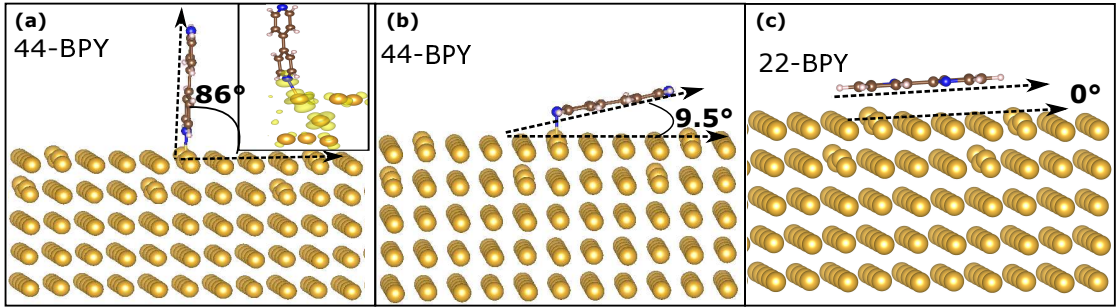}

\caption{Complete optimised structure for vertically aligned and horizontally aligned orientations of the molecules shown in panels (a) and (b) for $4,4^{\prime}$-BPY and in panel (c) for $2,2^{\prime}$-BPY.  Charge density of a state formed as a result of interactions between Au and the molecule (see SI Figure~S7) is shown in the inset of panel (a). } \label{figure5}
\end{figure*}

To shed more light on this, we have placed both molecules on gold surface and carried out static
calculations for different orientations of both molecules at a fixed distance. These results suggest
that there are two minima present for $4,4^{\prime}$-BPY on gold, while there is only one minima for
the $2,2^{\prime}$-BPY. Allowing for a complete structural optimization in each case leads to the
orientations of the molecules shown in panels (a) and (b) of Figure~5 for $4,4^{\prime}$-BPY and
what is shown in panel (c) for $2,2^{\prime}$-BPY. While in panel (a) we find that the molecule
prefers an orientation almost vertical to the surface, in (b) and (c) we find that the molecule
likes to lie horizontal to the surface. The reason for this is easy to understand. In the horizontal
position the molecule interacts through the carbon backbone with the Au surface. Hence for both
molecules, this corresponds to a stable orientation. In the vertical orientation, $4,4^{\prime}$-BPY
interacts strongly through the nitrogen with Au surface. This is evident from the charge density
shown in the inset of Figure~5a, where one finds significant weight on several Au atoms beyond it's
nearest neighbour. In contrast the Au-N bondlength is much longer for $2,2^{\prime}$-BPY and found
to be 2.24 $\AA$ as the steric effects from the neighbouring atoms don't allow the molecule to come
closer to the Au surface. This leads to only one stable minima for $2,2^{\prime}$-BPY. The strong
bonding interaction of the nitrogen in $4,4^{\prime}$-BPY that we find is extremely directional.
This would therefore imply that in the push cycle, we would not have have the desired
reproducibility as only certain orientations would form strong bonds. However, as we saw in the MD
simulations and also supported by an analysis of the charge density of the bonding state between
$4,4^{\prime}$-BPY and the Au surface, we find the interaction between them to be strong enough to
pull one or more Au atoms out. This then leads to a contact between two Au atoms in the push cycle,
needless to say the bonding between Au atoms involves s orbitals and is isotropic (it does not need
to attach back to its initial position, can attach with any other gold atom)
~\citep{Ruitenbeek_PRL_JUmptocontact,PhysRevLett.120.076802,PhysRevB.93.085437}, and hence we have
the reproducibility. From Figure~3c we observe that a jump to contact is observed in $\sim 20\%$ cases despite not forming any molecular junction during pull. As we found that $4,4^{\prime}$-BPY has two stable orientation on the Au surface, a jump may occur between them via a soft phonon mode of the molecule during push\citep{Wulfhekel_2008_PRB}.

\section{conclusion}
In conclusion, we have studied the formation and the post rupture evolution of molecular junctions
at room temperature of two isomers of bipyridene ($4,4^{\prime}$-BPY and $2,2^{\prime}$-BPY) having the same anchoring
groups using MCBJ techniques. Transport measurements indicate formation of a stable molecular junction during opening cycle for both of these molecules. During the closing cycle, however, we see that $4,4^{\prime}$-BPY shows clear formation of a
molecular junction via a jump in the conductance from the tunneling regime, while a continuous quantum tunneling like behavior was observed for $2,2^{\prime}$-BPY. As such formations of
molecular junctions via a jump in the conductance at room temperature are rare, we use DFT
calculations as well as ab-initio MD simulations to explain them. Two possible mechanisms have come out from our study. Firstly,  $4,4^{\prime}$-BPY was found to have two possible conformations on the Au surface due to the formation of strong Au-N bond, whereas, $2,2^{\prime}$-BPY prefers to lay flat on the Au surface. These two stable conformations of $4,4^{\prime}$-BPY may provide jump to contact while approaching the two electrodes. Secondly, we show that while breaking the
Au-$4,4^{\prime}$-BPY junctions, primarily the Au-Au bond that is ruptured instead of the molecule-Au bond. As
the Au atom attached to the molecule can bind isotropically with the other Au atoms of the
electrodes, one has single molecular junctions formed during the closing process via conductance
jumps similar to what one has seen for Au-Au contacts. Statistical analysis of our experimental data indicate that the second mechanism is more dominant one in our case. We provide an important insight, describing the role of metal-molecule interaction on the formation of chemical bond, important for the development of molecular-scale electronics.  

\section{Acknowledgement} 
A. N. Pal acknowledge the funding from Department of Science and Technology (grant no. CRG/2020/004208), B. Pabi acknowledges support from DST-Inspire fellowship, D. Mondal acknowledges support from CSIR-HRDG fellowship and P. Mahadevan acknowledges support from DST - Nanomission for the research through the project DST/NM/NS/2018/18.

%\bibliographystyle{pnas-new}
%\bibliographystyle{rsc}
%\bibliography{RefA9}
%\bibliography{RefA10}

\end{document}